\documentstyle[epsf]{elsart}  
\begin{document}
\newcommand{\p}{\partial}
\newcommand{\ls}{\left(}
\newcommand{\rs}{\right)}
\newcommand{\beq}{\begin{equation}}
\newcommand{\eeq}{\end{equation}}
\newcommand{\beqa}{\begin{eqnarray}}
\newcommand{\eeqa}{\end{eqnarray}}
\newcommand{\bdm}{\begin{displaymath}}
\newcommand{\edm}{\end{displaymath}}
%%%%%%%%%%%%%%%%%%%%%%%%%%%%%%%%%%%%%%%%%%%%%%%%%%%%%%%%%%%%%%%%%%%%%%%%%
%                                                                       %
%   BEGIN OF DOCUMENT                                                   %
%                                                                       %
%%%%%%%%%%%%%%%%%%%%%%%%%%%%%%%%%%%%%%%%%%%%%%%%%%%%%%%%%%%%%%%%%%%%%%%%%
\begin{frontmatter}
\title{
Kaon squeeze-out in heavy ion reactions
} 
\author{
Z. S. Wang, C. Fuchs, Amand Faessler and T. Gross-Boelting
}
\address{Institut f\"ur Theoretische Physik der 
Universit\"at T\"ubingen, D-72076 T\"ubingen, Germany}
%************************************************************************
\begin{abstract}
The squeeze-out phenomenon of $K^+$ and $K^-$ mesons, i.e. the
azimuthal asymmetry of $K^+$ and $K^-$ mesons emitted at 
midrapidity in heavy ion reactions,
is investigated for beam energies of 1-2 A.GeV. 
It is found that the squeeze-out
signal is strongly affected by in-medium potentials
of these mesons. The repulsive $K^+$-nucleus potential
gives rise to a pronounced
out-of-plane emission of $K^+$'s at midrapidity. With the $K^+$
potential we
reproduce well the experimental data of the $K^+$ azimuthal
distribution. It is found that the attractive $K^-$-nucleus potential
cancels to a large extent the influence of rescattering
and reabsorption of the $K^-$ mesons on the 
projectile and target residuals (i.e. shadowing).
This results in an azimuthally isotropic emission of the
midrapidity $K^-$ mesons with transverse momentum up to 0.8 GeV/c.
Since it is well accepted that the shadowing alone would
lead to a significant out-of-plane preference of particle emission, 
in particular
at high transverse momenta, the disappearance of the out-of-plane
preference for the $K^-$ mesons can serve as an unambiguous 
signal of the attractive
$K^-$ potential. We also apply a covariant 
formalism of the kaon dynamics
to the squeeze-out phenomenon. Discrepancies 
between the theory and the experiments and possible solutions are 
discussed.
\end{abstract}
\end{frontmatter}
%%%%%%%%%%%%%%%%%%%%%%%%%%%%%%%%%%%%%%%%%%%%%%%%%%%%%%%%%%%%%%%%%%%
\section{Introduction}
%%%%%%%%%%%%%%%%%%%%%%%%%%%%%%%%%%%%%%%%%%%%%%%%%%%%%%%%%%%%%%%%%%%
The asymmetry of particle emission with respect to the reaction plane
in heavy ion reactions has attracted increasing
interests of both experimentalists and theorists.
It has been found recently that the
azimuthal asymmetry of nucleons depends significantly on the bombarding
energy \cite{ollit98}. If the incident energy is comparable to the
Fermi energy, the attractive nuclear mean field dominates the dynamics.
Projectile and target form a rotating system
which preferentially emits nucleons in the reaction 
plane \cite{tsang84}. At high energies,
a positive pressure develops in the overlap region of the colliding nuclei
(so-called fireball) as a result of individual nucleon-nucleon 
collisions. The emission of the fireball nucleons can be
hindered by the less disturbed matter located in the reaction
plane (spectators). This shadowing leads to more nucleons emitted
in the direction out of the reaction plane than in the 
plane \cite{gutbr89,basti97}.
However, at ultra-relativistic energies the shadowing by spectator
matter is suppressed since the spectators depart too fast from the
fireball region. Then the nucleons can escape freely from the fireball.
Since the fireball has a smaller size along the impact parameter,
the pressure gradient is larger in the reaction plane. This results in
an in-plane preference of the nucleon emission \cite{barre94,sorge97}.
The transition from an in-plane preference due to rotation to an
out-of-plane one caused by shadowing was observed 
by the FOPI Collaboration to occur at an incident 
energy of 80-120 A.MeV \cite{basti97}.
The second transition to an in-plane preference at high energies 
takes place at 4-6 A.GeV \cite{ollit98,danie98}.

In this work we study the azimuthal asymmetry of $K^+$ and $K^-$ mesons
in heavy ion reactions at incident energies of 1-2 A.GeV. As
mentioned above, shadowing dominates these reactions and leads to
more nucleons out of the reaction
plane. The phenomenon of an out-of-plane
enhancement of particle emission at midrapidity is called "squeeze-out".
E.g., pions exhibit a clear out-of-plane preference 
\cite{brill93,venem93} which is due to shadowing, 
namely to scattering and absorption of the pions on the 
nucleons \cite{bli94,bass95}.
It is of high interest to see if the same mechanism still holds for kaons.
Unlike a pion, which keeps the mass more or less unchanged in a nuclear
environment, a kaon might vary its properties dramatically
in the medium due to chiral symmetry restoration \cite{kapla86}. 
The $K^+$- and $K^-$-nucleus potentials have observable effects
in heavy ion reactions: they can change the transverse flow of
kaons \cite{li95,wang97,bratk97,li98,cassing99}, the production
yields \cite{li98,cassing99,li97}, and may give rise to a 
collective flow, i.e. a radial flow \cite{wang98}.

This paper is organised as follows: In section 2 we describe the kaon
dynamics within the framework of the
Quantum Molecular Dynamics (QMD) model. The kaon potential is based on 
chiral perturbation theory, however, with
the space-like components of the baryon current neglected as usual. 
We present the azimuthal asymmetry of $K^+$ mesons 
in section 3, and the corresponding results for $K^-$ 
in section 4. Section 5 contains a detailed discussion of a covariant
treatment of the in-medium kaon dynamics. In section 6 the paper is
summarised. 
%%%%%%%%%%%%%%%%%%%%%%%%%%%%%%%%%%%%%%%%%%%%%%%%%%%%%%%%%%%%%%%%%%%%%%%%%
\section{Kaon dynamics in heavy ion reactions}
%%%%%%%%%%%%%%%%%%%%%%%%%%%%%%%%%%%%%%%%%%%%%%%%%%%%%%%%%%%%%%%%%%%%%%%%%
We adopt the QMD model to describe the dynamics of heavy
ion reactions. $K^+$ and $K^-$ mesons are produced from baryon-baryon
collisions or pion-baryon collisions. The corresponding cross sections 
for the $K^+$ production are taken from 
Refs. \cite{sibirtsev,tsushima} for 
the baryon-baryon channels and from \cite{tuebingen} for the pion-baryon 
induced channels. The $K^-$ production is treated as described in 
Ref. \cite{cassing99} with the corresponding elementary cross sections 
given there. The present QMD model describes successfully a 
large set of observables, 
such as production cross sections and collective flow 
patterns of nucleons, pions and kaons \cite{uma98,fuchs97,wang972}.
In particular, this model
describes well the experimental transverse flow of protons,
$K^+$ mesons as well as $\Lambda$ hyperons \cite{wang97,wang982}. 

While a $K^+$ meson is hardly
absorbed in nuclear matter due to strangeness conservation, 
a $K^-$ meson can be easily annihilated through the 
reaction $K^-$ + N $\rightarrow$
$\pi$ + Y, where Y denotes a $\Sigma$ or a $\Lambda$ hyperon.
Both $K^+$ and $K^-$ mesons can scatter elastically with nucleons.
The $K^-$ absorption cross section is
larger than 50 mb for $K^-$ momenta below 0.2 GeV/c. 
Compared to this value the elastic  $K^+ N$ cross section 
is relatively small ($\sigma_{K^+N}$ $\approx$ 10 mb) which results in a 
long mean free path of $K^+$ mesons in nuclear matter. 
In addition to scattering and absorption process, the strong interaction
of $K^+$ and $K^-$ mesons with a nuclear medium gives rise
to a mean field which acts on the kaons when they propagate
in the medium. Here both, the strong and the Coulomb potential 
are included \cite{wang98}. 

The $K^+$ and $K^-$ potentials are defined in the usual way as the
difference of the in-medium dispersion relation and the free one
\beq
U_K = \omega_K - \sqrt{m_K^2 + \vec{p}^2}.
\eeq
Starting from an $SU(3)_L$$\times$$SU(3)_R$ chiral Lagrangian as proposed by
Kaplan and Nelson \cite{kapla86}, one obtains the field equations for $K^+$
and $K^-$ mesons in mean field approximation \cite{li95},
\beq
[ \partial_{\mu}\partial^{\mu} \pm \frac{3i}{4f^2_{\pi}}j_{\mu}
\partial^{\mu} + ( m^2_K - \frac{\Sigma_{KN}}{f^2_{\pi}}\rho_s )]
\phi_{K^{\pm}}(x) = 0 \quad ,
\eeq
where $\rho_s$ is the scalar baryon density and $j_{\mu}$ the baryon 
four-vector current. $f_{\pi}$ $\approx$ 93 MeV is the 
pion decay constant. According to recent lattice QCD
calculations \cite{brown96} the kaon-nucleon sigma term 
is taken to be $\Sigma_{KN}$ $\approx$ 450 MeV. 
The term coupling to the baryon current
(so-called Weinberg-Tomozawa term) is of leading order
in the chiral expansion, while the sigma term (so-called Kaplan-Nelson term)
comes from the next order. The in-medium dispersion 
relation reads \cite{li95,fuchs98}
\beq
\omega^2 = m^2_K + \vec{p}^2 - \frac{\Sigma_{KN}}{f^2_{\pi}}\rho_s
+ \frac{3}{4f^2_{\pi}}(\rho_B\omega - \vec{j} \cdot \vec{p}).
\eeq
Usually the spatial components $\vec{j}$ of the baryon current are 
neglected \cite{li95,wang97,bratk97,li98,cassing99,li97}. This means 
that the kaons are propagated in a static, momentum independent 
potential. Since the resulting kaon dynamics
has been shown to agree well with experiments 
concerning, e.g., the transverse flow \cite{li95,wang97,li97,li98} 
we follow this conventional treatment.
However, one looses Lorentz covariance by the neglection 
of the spatial components of the four-vector baryon current 
\cite{fuchs98}. Thus, in Section 5 we will also discuss
the covariant treatment of the kaon dynamics which includes 
the momentum dependence of the mean field in lowest order, i.e. 
that part which arises from Lorentz boosts.

Brown and Rho pointed out \cite{brown962} that the range 
term, which is of the same order in the
chiral expansion as the $\Sigma_{KN}$ term also
contributes to the mean field. Moreover, the pion decay constant
$f_{\pi}$ can be reduced in the medium due to the decreasing 
quark condensate. For $K^+$ the range term and the medium modification of
the pion decay constant are taken into account. 
At nuclear matter saturation density $\rho_0$ = 0.16 $fm^{-3}$ 
this potential agrees with the empirical knowledge obtained 
from the impulse approximation to free $K^+$N scattering data 
\cite{barne94}. The $K^-$ potential is similar to
that used in other works \cite{li95,bratk97,cassing99,li97}.
It is constructed according to eq.(1) and eq.(3), 
however, with a smaller value of $\Sigma_{KN}$ = 350 MeV and
the $f_{\pi}$ taken as in free space. The range term is neglected.
This potential is less attractive than
that extracted from kaonic atoms: the former is about -100 MeV
at $\rho_0$ while the latter is about -185 $\pm$ 15 MeV 
\cite{mille88}. However, such a less attractive $K^-$ 
potential seems to be necessary in order to reproduce
the experimental $K^-$ yield in heavy ion reactions at
1-2 A.GeV \cite{cassing99,li97}. This inconsistency is so far unsolved 
(see also Section 5). 
In Fig.1 we show the potentials for $K^+$ and $K^-$
at zero momentum as a function of the nuclear matter density. 

In addition to the influence on the propagation, 
the potentials can also change the production thresholds of the mesons 
and thus the corresponding yields.
For $K^-$ the potential has been included in the threshold 
throughout the present work in the same way as in Refs. 
\cite{cassing99,li97}. The influence of the $K^+$ in-medium 
potential on the yields is demonstrated in Fig.2. The reaction 
considered is $C+C$ at 2.0 A.GeV under minimal bias conditions. 
For the comparison to the corresponding KaoS data \cite{kaos99} 
we applied a $\Theta_{\rm Lab} = 40^o \pm 4^o $ polar angular cut. 
We distinguish three different cases: First a calculation without 
any medium effects, a calculation where the potential is only 
included in the propagation and finally a full 
calculation where the potential is included in the threshold 
as well as in the propagation. Without medium effects we are able 
to reproduce the low $p_t$-region of the spectrum 
but underpredict the high $p_t$ part. On the other hand, the potential 
effect on the thresholds strongly suppresses the low energetic kaons. 
If the potential also acts on the propagation it makes the spectrum 
harder which is due to the repulsive forces. This can be clearly seen 
from the calculation where the potential is only included the 
propagation of the kaons. However, the uncertainty 
in the description of the cross section 
is still too large in order to draw definite conclusions on 
the importance of medium effects. Although 
our calculation is in reasonable agreement with the results 
of other groups \cite{li97,cassing99}, there remain still 
discrepancies which reflect the uncertainties in the theoretical 
knowledge of the elementary production cross sections. 
Thus, for $K^+$ we will neglect the effect of the potential on the 
production thresholds in the following. This treatment is also 
justified since we study collective flow phenomena which do not 
depend sensitive on total yields. 
%%%%%%%%%%%%%%%%%%%%%%%%%%%%%%%%%%%%%%%%%%%%%%%%%%%%%%%%%%%%%%%%%%%%%%%%%
\section{$K^+$ squeeze-out}
%%%%%%%%%%%%%%%%%%%%%%%%%%%%%%%%%%%%%%%%%%%%%%%%%%%%%%%%%%%%%%%%%%%%%%%%%
The strength of the azimuthal asymmetry can be quantified by the ratio
of the particle multiplicity emitted perpendicular to the reaction plane
over the multiplicity parallel to the plane
\beq
R_{out/in} = \frac{N(\phi=90^0) + N(\phi=270^0)}{N(\phi=0^0) + N(\phi=180^0)}.
\eeq
The azimuthal angles $\phi$ = $0^0$ and $180^0$ correspond to the positions
of target and projectile in the reaction plane, while $\phi$ = $\pm$$90^0$
denote the directions perpendicular to the plane.
A ratio $R_{out/in}$$>$1 means a preference of the particle emission
out of the reaction plane. An azimuthal distribution can be expressed
in terms of a Fourier series \cite{volos97}
\beq
\frac{dN}{d\phi} \sim C[ 1 + a_1cos(\phi) 
+ a_2cos(2\phi) + a_3cos(3\phi) + ...].
\eeq
The dipole term arises from a collective sideward deflection of 
the particles in the reaction plane ("transverse flow") 
and does not contribute to the squeeze-out ratio $R_{out/in}$. In 
a symmetric collision, the cos(3$\phi$) term vanishes. The squeeze-out ratio
is then determined only by the quadrupole component
\beq
R_{out/in} = \frac{1 - a_2}{1 + a_2}.
\eeq
The azimuthal asymmetry of the $K^+$ production in heavy ion 
reactions has been first studied by Li et al. \cite{li96}. 
However, in \cite{li96} the authors focused on 
kaons emitted at projectile or target rapidity.
In the present study we will focus on midrapidity kaons which 
should yield more precise information on the dense fireball. 

In Fig.3 the $K^+$ multiplicity as a function of azimuthal
angle $\phi$ is shown for a semi-central (b=6 fm) $Au + Au$ 
reaction at  1 A.GeV incident energy. A transverse momentum cut of $P_T$
$>$ 0.2 GeV/c and a rapidity cut centred at midrapidity, i.e.
-0.2 $<$ $(Y/Y_{proj})^{cm}$ $<$ 0.2, have been applied. 
The same cuts have been adopted by the KaoS Collaboration. 
The KaoS data \cite{shin98} for the same reaction at 
semi-central impact parameters (corresponding to 5 fm $<$ b $<$ 10 fm)
are also shown. We performed the QMD calculations for three
different scenarios: 

(a) with full in-medium $K^+$ dynamics;\\
(b) the $K^+$ potential due to the strong interaction is neglected;\\
(c) in addition, the Coulomb potential is neglected.

It can be seen from Fig.3 that the full calculation leads to
an enhanced out-of-plane ($\phi$ = $90^0, 270^0$) 
$K^+$ emission. The corresponding ratio $R_{out/in}$ = 1.5 
is close to the experimental
value of $R_{out/in}$ = 1.7 \cite{shin98}. We would like to mention that
we also performed calculations for different impact parameters 
ranging from b=3 to b=10 fm. From the impact parameter dependence of
the kaon production we found that it is reasonable to compare 
the simulation at the single impact parameter b=6 fm 
with the experimental data since the kaon multiplicity 
decreases very fast with decreasing centrality. (A corresponding 
calculation of the Stony Brook group \cite{shin98} has been 
performed for the representative impact parameter b=7 fm.)

Comparing the cases (b) and (c) one clearly sees 
that the $K^+$ out-of-plane preference
is mainly a result of the strong potential. The $K^+$ emission is
nearly azimuthally isotropic in the calculation with neither the strong
nor the Coulomb potential, while the Coulomb potential slightly increases
the out-of-plane abundance. However, the Coulomb force has only a minor 
effect, and leads to an azimuthal asymmetry which is much 
weaker than the experimental one. Generally our results are in good 
agreement with those found by the Stony Brook group \cite{shin98}. 

Fig.4 shows the $K^+$ $R_{out/in}$ ratio at midrapidity
as a function of the transverse momentum. In this figure we 
also show a calculation where all
final-state interactions including $K^+ N$ rescattering, 
have been neglected. The primordial $K^+$ mesons 
exhibit a slight in-plane preference
which increases with the momentum. This behaviour reflects the
primordial in-plane $K^+$ flow which follows the flow pattern of the
production sources, i.e. the in-plane flow of the nucleons. 
Rescattering enhances the out-of-plane emission of the 
$K^+$ mesons and leads to a nearly isotropic azimuthal distribution. 
Thus we observe a clear shadowing effect by the spectator matter. 
However, the effect of $K^+$-nucleon scattering is much less pronounced
than that of the strong potential. The strong potential 
enhances dramatically the out-of-plane emission
at transverse momentum between 0.2 GeV/c and 0.6 GeV/c. Now 
the azimuthal asymmetry exhibits a complex momentum 
dependence since the out-of-plane preference decreases 
again at high momenta. 

This momentum dependence is, however, understandable.
First of all, midrapidity $K^+$ mesons experience the strongest repulsion
in the dense fireball. The repulsive potential gradient
is larger in the direction perpendicular to the reaction plane 
than parallel to the plane. 
In the configuration where the fireball is combined with the spectators
the distance from the fireball centre to free space is much shorter
along the out-of-plane direction than in the reaction plane.
Consequently, the $K^+$ mesons are driven by the potential 
gradient preferentially out of plane. With other words, in the reaction 
plane the kaons are repelled by the spectator matter. Thus, the 
potential acts similar as the elastic scattering and strongly enhances 
the shadowing effect. There occurs, however, a difference 
between the effect of the potential and the shadowing by simple 
rescattering/absorption on spectator matter. From nucleons and pions 
\cite{basti97,brill93,venem93} it is known that the shadowing 
enhances the out-of-plane emission of high energetic particles, i.e. 
one observes an increasing $R_{out/in}$ ratio with increasing 
momentum. The same effect is observed in our calculation for 
$K^-$ discussed in the next section. This is understandable 
since in these cases the high energetic particles stem mostly from the 
early phase of the reaction where the fireball and the spectators 
are clearly developed. However, the $K^+$ squeeze-out shows a different 
trend. Recent data \cite{shin98} indicate a constant $R_{out/in}$ 
ratio whereas the calculation including the $K^+$ potential shows 
even a decreasing $R_{out/in}$  with increasing momentum. 
To understand this behaviour one 
has to keep in mind that the repulsive $K^+$ potential accelerates the 
particles and makes the spectrum harder (see also Fig.2). 
High energetic $K^+$ mesons which experience the acceleration by 
the medium for a longer time span stem probably from the 
later stages of the reaction. But this means that the fireball and the 
spectators are washed out to more extent which results in less 
shadowing and a more isotropic azimuthal distribution of the 
particles. However, it appears that for a 
complete understanding of the $K^+$ 
squeeze-out phenomenon, also its momentum impact parameter dependence, 
further going studies seem to be necessary. 
%%%%%%%%%%%%%%%%%%%%%%%%%%%%%%%%%%%%%%%%%%%%%%%%%%%%%%%%%%%%%%%%%%%%%%%%%
\section{$K^-$ squeeze-out}
%%%%%%%%%%%%%%%%%%%%%%%%%%%%%%%%%%%%%%%%%%%%%%%%%%%%%%%%%%%%%%%%%%%%%%%%%
$K^-$ mesons are strongly scattered or absorbed in the nuclear medium.
In the absence of an additional in-medium potential, 
one expects a strong out-of-plane emission of midrapidity $K^-$ mesons 
much like pions, since in both cases shadowing
by the spectators plays a dominant role.
In Fig.5 we present the azimuthal distribution of the $K^-$ mesons emitted at
midrapidity (-0.2 $<$ $(Y/Y_{proj})^{cm}$ $<$ 0.2) in the
reaction of $Au+Au$ at E/A = 1.8 A.GeV and b=8 fm.
A $P_T$ cut of $P_T$ $>$ 0.5 GeV/c has been used. Fig.6 shows the 
corresponding $R_{out/in}$ as a function of the 
transverse momentum. Let us first consider the calculations \
where the strong and Coulomb potentials are neglected.
Figs.5 and 6 show that the $K^-$ mesons are then 
preferentially emitted out of the reaction
plane. The $R_{out/in}$ ratio increases with transverse momentum.
This $P_T$ dependence is very similar to that observed experimentally
for nucleons and pions \cite{basti97,brill93,venem93}.

In Figs.5 and 6 the results of the full calculation
are presented as well. It can be seen that the $K^-$
in-medium potential reduces dramatically the out-of-plane abundance
($\phi$ = $90^0$ and $270^0$), and leads thereby to a nearly
isotropic azimuthal emission. The $R_{out/in}$ ratio remains 
more or less flat as a function of the transverse momentum. 
We can understand the effect of the in-medium
potential acting on $K^-$ in a similar way as we have done for $K^+$.
The $K^-$ potential is more attractive
at higher densities. Consequently, the $K^-$ mesons feel
an attractive potential gradient as they move from the dense fireball
to free space. This potential gradient is larger in the out-of-plane direction
than in the in-plane direction when the fireball is still connected
with the spectators. Thus the potential now favours the $K^-$
emission in the reaction plane.

It is found in our calculations that the Coulomb potential
has the same tendency but with a much smaller magnitude.
From previous studies on nucleons and pions,
both experimental \cite{basti97,brill93,venem93} 
and theoretical \cite{bli94,bass95}, it is already clear that
frequent rescattering and re-absorptions will lead to a pronounced
out-of-plane preference of particle emission and an 
increasing squeeze-out ratio with increasing transverse momentum.
Therefore, the disappearance of the out-of-plane preference 
up to transverse momentum of $P_t$ = 0.8 GeV/c
can serve as an unambiguous signal for the 
attractive $K^-$ in-medium potential.

So far we have shown that the azimuthal asymmetry of $K^+$
and $K^-$ mesons in heavy ion reactions at 1-2 A.GeV is
sensitive to the in-medium potentials.
The considered beam energies justify the mechanism where shadowing,
namely rescattering and reabsorption by the spectators,
compete with the in-medium potentials in the evolution of an
kaon azimuthal asymmetry. In order to minimise the shadowing 
effect shadowing, one can study higher beam energies,
i.e. above 6 A.GeV, where
an isolated fireball elongated
in the direction perpendicular to the reaction plane, rather than
a combination of the fireball and the spectators, is responsible for
the azimuthal asymmetry of particle emission at midrapidity.
At such high beam energies, shadowing plays a minor role, and therefore,
the azimuthal asymmetry of the midrapidity kaons yields a cleaner information
on the in-medium potentials.
%%%%%%%%%%%%%%%%%%%%%%%%%%%%%%%%%%%%%%%%%%%%%%%%%%%%%%%%%%%%%%%%%%%%%%%%%
\section{Covariant kaon dynamics}
%%%%%%%%%%%%%%%%%%%%%%%%%%%%%%%%%%%%%%%%%%%%%%%%%%%%%%%%%%%%%%%%%%%%%%%%%
We have mentioned that the kaon dynamics is described in the present study
in a non-covariant way since the spatial components of the baryon current
are neglected. Such a description is correct in nuclear matter at rest
but generally not in energetic heavy ion collisions.
Keeping the four-vector baryon current in the kaon field equation,
eq.(2), one can construct a covariant description of the kaon
dynamics in the nuclear medium \cite{fuchs98}. In order to do this,
it is convenient to write the kaonic vector potential as
\beq
V_\mu = \frac{3}{8f_\pi^2}j_\mu 
\quad ,
\eeq
The effective mass of the kaons is defined by
\beq
m_k^* = \sqrt{ m_k^2 - \frac{\Sigma_{KN}}{f_\pi^2}\rho_s + V_{\mu}V^{\mu} } 
\quad .
\eeq
The kaon field equation is then rewritten as
\beq
[ (\partial_{\mu} + iV_{\mu})^2 + m_k^* ] \phi_{K^+} (x) = 0,
\eeq
\beq
[ (\partial_{\mu} - iV_{\mu})^2 + m_k^* ] \phi_{K^-} (x) = 0.
\eeq
Thus the vector field is introduced by minimal coupling into
the Klein-Gordon equation. The effective mass defined by eq.(8) is
a Lorentz scalar and is equal for $K^+$ and $K^-$ mesons \cite{fuchs98}.
We should note that this definition of kaon effective mass is different
from that conventionally used in the non-covariant description
of the kaon dynamics. In the latter case the effective mass of the kaons
is defined as kaon in-medium energy at zero momentum. Introducing
an effective kinetic momentum
\beq
p_{\mu}^* = p_{\mu} \mp V_\mu
\eeq
for $K^+$ ($K^-$) mesons, the Klein-Gordon equations (eq.(9) and eq.(10))
reads in momentum space
\beq
[ {p^*}^2 - {m^*}^2 ] \phi_{K^{\pm}} (p) = 0.
\eeq
This is just the mass-shell condition for the quasi-particles carrying
the effective mass and kinetic momentum, i.e. the effective energy is
on-shell $E^*$ = $p_0^*$ = $\sqrt{ \vec{p^*}^2 + {m^*} ^2 }$.
These quasi-particles
can be treated like free particles. This yields in the test-particle
limit the covariant equations of motion for the kaons \cite{fuchs98}
\beqa
\frac{d \vec{q}}{dt} &=& \frac{\vec{p^*}}{E^*} \\
\frac{d \vec{p^*}}{dt} &=& - \frac{m_k^*}{E^*} \frac{\partial m_k^*}
{\partial \vec{q}} \mp \frac{\partial V^0}{\partial \vec{q}}
\mp \partial_t \vec{V}
\pm \frac{\vec{p^*}}{E^*}\times (\frac{\partial}{\partial \vec{q}}
\times \vec{V})
\eeqa
where the upper (lower) signs refer to $K^+$ ($K^-$) mesons. The last
term of eq.(14) provides a momentum dependent
force which is missing in the non-covariant description of the kaon dynamics.
Such a force is analogous to the Lorentz
force in electrodynamics, and is a genuine feature of relativistic
dynamics as soon as a vector field is involved. Since in the QMD
approach the equations of motion are formulated in terms of canonical
momenta $\vec{p}$ rather than kinetic momenta $\vec{p^*}$, it is
instructive to rewrite eq.(14)
\beq
\frac{d \vec{p}}{dt} = - \frac{m_k^*}{E^*} \frac{\partial m_k^*}
{\partial \vec{q}} \mp \frac{\partial V^0}{\partial \vec{q}}
\pm \vec{\beta} \cdot \frac{\partial \vec{V}}{\partial \vec{q}}
= -\frac{\partial}{\partial \vec{q}}U_K \pm \vec{\beta} \cdot \frac{
\partial \vec{V}}{\partial \vec{q}} 
\quad ,
\eeq
with $\vec{\beta}$ = $\vec{p}^* / E^*$ the kaon velocity and
the potential $U_K$ given by eq.(1). Eq.(15) is the equation which
we have solved in the present work. 
A study with use of a relativistic version of
the QMD model (RQMD) has shown that the inclusion of the 
Lorentz-like force leads
for the reaction Ni+Ni at 1.93 A.GeV $(b\leq 4$fm) to a $K^+$ 
transverse flow which essentially follows the nucleon flow \cite{fuchs98}.
This result contradicts the experimental data \cite{ritma95},
since the latter show clearly different flow patterns for 
nucleons and $K^+$. The QMD model used in the present 
work yields a result similar to that of the covariant RQMD
model after the Lorentz-like force is included. 
However, as shown in our previous work, the same QMD
model but in absence of the Lorentz-force enabled us 
to reproduce very well the FOPI data \cite{ritma95} 
of the in-plane $K^+$ flow. Other groups also got
agreement with the data using conventional kaon dynamics 
\cite{li95,bratk97}. 

In Fig.7 we show the azimuthal $K^+$ distribution
for the same reaction as in Fig.3, however, now including 
the Lorentz-force like momentum dependence. 
One finds again that this force destroys the agreement with 
experiments: the calculation with the momentum dependent force yields
an isotropic $K^+$ emission with respect to the azimuthal angle $\phi$,
while the KaoS data show a remarkable out-of-plane preference.

Both, the in-plane flow and the azimuthal asymmetry are predominately
determined by the interactions of the kaons with the spectators. Due
to the non trivial relative motion between the kaons and the spectators,
the momentum dependent Lorentz-force plays a role in cancelling 
the effect from the time-like component of the vector potential 
which is the major source of the repulsion acting on the  
$K^+$ mesons. Therefore we observe a $K^+$ in-plane flow
similar to the nucleon flow and an azimuthal isotropy of the 
midrapidity $K^+$ mesons. In a recent publication \cite{wang98} 
we demonstrated that, in heavy ion reactions
at a beam energy below the kaon production threshold in free 
space (1.58 A.GeV for $K^+$ and 2.5 A.GeV for $K^-$ mesons),
the in-medium potentials lead to new collective motion in the radial
direction of midrapidity $K^+$ and $K^-$ mesons. 
This collective flow results in a "shoulder-arm" or a "concave"
structure in the transverse mass spectrum of the  $K^+$ or  $K^-$ 
mesons, respectively.
Such "shoulder-arm" or "concave" structure obviously
differentiates the kaon spectrum from the standard Boltzmann
distribution. We called this collective motion kaon radial flow.
The kaon radial flow is primarily a consequence of 
the interactions of the midrapidity
kaons with the fireball. Since the relative
motion between the kaons and the fireball is small, the radial flow
has been found to keep more or less unchanged after the Lorentz-like 
force is included \cite{wangd}.

It is surprising that the conventional description of the in-medium kaon
dynamics, rather than the more consistent one including the full 
four-vector baryon current, turns out to agree 
with the phenomenology of kaon production in heavy ion
reactions. To understand this fact it is important
to realize that the kaon field equation, eq.(2), which is the same starting
point for the two treatments, consists only of the
two lowest-order contributions from the chiral expansion,
namely the Weinberg-Tomozawa and Kaplan-Nelson term.
At that level the mean field approximation gives rise to
only density dependent scalar and vector potentials for the kaons.
Although the consistent treatment is based on the same level, it
accounts for the additional momentum dependence which arises by
Lorentz covariance and thus addresses the momentum dependence of
the interaction of the kaons in the nuclear medium at lowest order.
The success of the conventional treatment where the 
momentum dependence is neglected
seems to imply that higher order contributions
in the chiral expansion might lead to
cancellation effects.
Thus, one probably needs to take into account
in a covariant formalism not only the lowest-order terms but also higher-order
contributions, e.g. P-wave contributions of the kaon-nucleus interaction 
arising from nucleon hole-hyperon
excitations. Such a P-wave interaction goes 
beyond the Weinberg-Tomozawa and Kaplan-Nelson terms 
and leads to a non trivial momentum dependence of the kaon 
potential \cite{kolom95,lutz97,sibir98}.

Furthermore, the mean field approximation which is used to obtain 
the field equations (2) is justified at 
low nuclear matter densities, but might be
questionable at high densities. The kaon-nucleon and 
nucleon-nucleon correlations will 
start to play a more important role \cite{pandh95}. 
The effect of the correlations can be
illustrated considering the low and high density limits. At very low
densities a kaon interacts many times with the same nucleon
before it encounters another one. Thus, the impulse approximation
is justified and the energy of a kaon interacting with nucleons is given
in terms of the kaon-nucleon scattering length $a_{KN}$ \cite{lutz97},
\beq
\omega_{Lenz} = m_K - \frac{2\pi}{m_K}(1+\frac{m_K}{m_N})a_{KN}\rho_B 
\quad .
\eeq
This is the so-called Lenz potential. At high densities, 
however, the K-N interaction has to be summed over 
many nucleons which results in a potential of a Hartree type. 
According to an estimation given in Ref. \cite{pandh95},
the Hartree potential is reduced compared to the Lenz potential
by a factor of 1.63 for $K^-$'s in nuclear matter.
The correlation effect for $K^+$'s should be smaller
than for $K^-$'s, since the $K^+$-nucleon interaction
is relatively weak compared to other hadron-nucleon interactions.
However, the onset of short-range correlations
is a general feature of the dynamics of dense matter. Such effects
have, e.g., been discussed concerning particle production in heavy
ion reactions \cite{wang95}. 
In Ref. \cite{heise98} the transition from the Lenz 
to the Hartree potential was investigated for
$K^-$ mesons in neutron matter. There it is found that correlations
reduce the $K^-$ potential already significantly at 
densities below 3$\rho_0$ which could also provide a solution
of the disagreement between the $K^-$ potential extracted from
kaonic atoms (at $\rho \leq \rho_0$) 
and the less attractive potential which has to be 
used to obtain a reasonable description of the
experimental $K^-$ yields in heavy ion reactions
at 1-2 A.GeV \cite{li98} (at $\rho \leq 3\rho_0$). 
%%%%%%%%%%%%%%%%%%%%%%%%%%%%%%%%%%%%%%%%%%%%%%%%%%%%%%%%%%%%%%%%%%%%%%%%%
\section{Conclusions}
%%%%%%%%%%%%%%%%%%%%%%%%%%%%%%%%%%%%%%%%%%%%%%%%%%%%%%%%%%%%%%%%%%%%%%%%%
In the present paper we investigated the azimuthal
asymmetry of midrapidity $K^+$ and $K^-$ mesons in
heavy ion reactions at beam energies of 1-2 A.GeV.
The in-medium kaon potential is constructed from the lowest-order
terms of the chiral expansion. The kaon dynamics
are described in a conventional way with a quasi-potential formalism where
the space-like components of the baryon current are neglected. This 
description turned out to be rather successful in reproducing the 
current experimental data on $K^+$ and $K^-$ production.

The present study finds that
the in-medium potentials of the $K^+$ and $K^-$ mesons play an important
role in the evolution of an azimuthal asymmetry. Due to the long mean free
path of the $K^+$ mesons in nuclear matter, $K^+$-nucleon scattering
processes are insufficient to understand the 
pronounced out-of-plane emission of
midrapidity $K^+$ mesons observed by the KaoS Collaboration.
The $K^+$ potential drives the kaons out of the reaction plane
by a repulsive potential gradient. These kaons are repelled by the 
spectators which leads to an additional shadowing. The repulsive 
potential is necessary in order to understand the KaoS data. The 
momentum dependence of the $K^+$ squeeze-out signal is, however, 
more complex. 
We observe a decreasing squeeze-out signal with increasing transverse 
momentum. For a complete understanding of the $p_t$ 
dependence of the $K^+$ squeeze-out further-going theoretical 
and experimental investigations seem to be necessary.

The $K^-$ potential, in contrast, suppresses
the $K^-$ emission out of the reaction plane
by an attractive potential. The effect is found to
counterbalance to a large extent the strong 
shadowing by scattering and absorption.
This results in a nearly isotropic azimuthal $K^-$ emission. 
Since it is clear from both experimental and theoretical
studies that frequent scattering and absorption give rise to a remarkable
out-of-plane preference of the particle emissions, the disappearance of 
the squeeze-out can serve as a signal of
the $K^-$ in-medium potential. 

In a covariant treatment, in particular with respect to the 
vector potential which is proportional to the 
four-vector baryon current, one finds that the agreement 
with experiments concerning both, the 
azimuthal asymmetry and the in-plane flow is destroyed. 
Since the corresponding fields originate from the lowest-order 
terms of the kaon-nucleon interaction we suggest to 
include in future also higher-order contributions, in 
particular to treat the momentum dependence of the kaon-nucleon 
interaction more carefully.

%%%%%%%%%%%%%%%%%%%%%%%%%%%%%%%%%%%%%%%%%%%%%%%%%%%%%%%%%%%%%%%%%%%%%%%%%
{\bf Acknowledgments}
%%%%%%%%%%%%%%%%%%%%%%%%%%%%%%%%%%%%%%%%%%%%%%%%%%%%%%%%%%%%%%%%%%%%%%%%%
The authors would like to thank M. Lutz for valuable discussions. 
%%%%%%%%%%%%%%%%%%%%%%%%%%%%%%%%%%%%%%%%%%%%%%%%%%%%%%%%%%%%%%%%%%%%%%%%%
%   END OF TEXT                                                         %
%                                                                       %
%%%%%%%%%%%%%%%%%%%%%%%%%%%%%%%%%%%%%%%%%%%%%%%%%%%%%%%%%%%%%%%%%%%%%%%%%
%%%%%%%%%%%%%%%%%%%%%%%%%%%%%%%%%%%%%%%%%%%%%%%%%%%%%%%%%%%%%%%%%%%%%

%%%%%%%%%%%%%%%%%%%%%%%%%%%%%%%%%%%%%%%%%%%%%%%%%%%%%%%%%%%%%%%%%%%%%%%%%
%                                                                       %
%   END OF THEBILIOGRAPHY                                               %
%                                                                       %
%%%%%%%%%%%%%%%%%%%%%%%%%%%%%%%%%%%%%%%%%%%%%%%%%%%%%%%%%%%%%%%%%%%%%%%%%
%%%%%%%%%%%%%%%%%%%%%%%%%%%%%%%%%%%%%%%%%%%%%%%%%%%%%%%%%%%%%%%%%%%%%%%%%
%                                                                       %
%   BEGIN OF FIGURES                                                    %
%                                                                       %
%%%%%%%%%%%%%%%%%%%%%%%%%%%%%%%%%%%%%%%%%%%%%%%%%%%%%%%%%%%%%%%%%%%%%%%%%
\newpage
{\bf Figure Captions:}
\vskip 0.5 true cm

Fig. 1. The kaon potential
in nuclear matter used in this work (at zero momentum). The
filled circle shows the potential for $K^-$ mesons at normal nuclear matter
density $\rho_0$ extracted from kaonic atoms \cite{mille88},
while the open circle presents the potential for $K^+$ mesons obtained 
from the impulse approximation to free $K^+$-nucleon scattering data 
\cite{barne94}.

\vskip 0.5 true cm

Fig. 2. Invariant cross section of the $K^+$ production 
in C+C collisions at 2 A.GeV. The The calculations are performed 
without any medium effects (dotted), with full medium effects (solid), 
and including the in-medium potential only for the propagation of 
the kaons but neglecting it in the production thresholds 
(long-dashed). A $\Theta_{\rm lab} = 40\pm 4^{o}$ polar angular cut 
has been applied in order to compare to the corresponding 
KaoS data taken from Ref. \cite{kaos99}.

\vskip 0.5 true cm

Fig. 3. The $K^+$ multiplicity as a function of the azimuthal angle
calculated with the QMD model for the Au+Au reaction at an incident
energy of 1 A.GeV and at an impact parameter of b = 6 fm.
The dotted line shows the result of the QMD calculation with the full
$K^+$ dynamics. The dashed line stands for
the calculation where the strong $K^+$ potential 
is neglected, while the dashed-dotted line refers to the calculation
where the Coulomb potential is
neglected as well. 
The KaoS data \cite{shin98} (solid line) for the same reaction 
at semi-central impact parameters 
(5fm$< b <$10fm) are also shown.
\vskip 0.5 true cm

Fig. 4. The $R_{out/in}$ ratio defined 
in eq.(4) for the $K^+$'s as a function of
transverse momentum for the same reaction as in Fig.3.
The solid line shows the result of the QMD calculation with the full
$K^+$ dynamics as described in text. The dashed line stands for
the calculation where the $K^+$ potentials of both the strong interaction
and the Coulomb interaction are neglected, while the dashed-dotted line refers
to the calculation
where in addition to the potentials $K^+$-nucleon scattering has
been neglected.

\vskip 0.5 true cm
Fig. 5. The $K^-$ multiplicity as a function of the azimuthal angle
calculated with the QMD model for the Au+Au reaction at an incident
energy of 1.8 A.GeV and at an impact parameter of b = 8 fm.
The solid line shows the result of the QMD calculation with the full
$K^-$ dynamics as described in text, while the dashed line stands for
the calculation where the $K^-$ strong and Coulomb potentials are
neglected.

\vskip 0.5 true cm
Fig. 6. The $R_{out/in}$ ratio, eq.(4), for $K^-$'s as a function of
transverse momentum for the same reaction as in Fig.5.
The solid line shows the result of the QMD calculation with the full
$K^-$ dynamics, while the dashed line stands for
the calculation where the $K^-$ strong and Coulomb potentials are
neglected.

\vskip 0.5 true cm
Fig. 7. The $K^+$ multiplicity as a function of the azimuthal angle
for the same reaction as in Fig.3.
The dotted line shows the result of the QMD calculation with the
$K^+$ dynamics using the conventional static potential, 
while the dashed line stands for
the calculation where the Lorentz-like force due to
the space-like components of the baryon current is included.
%%%%%%%%%%%%%%%%%%%%%%%%%%%%%%%%%%%%%%%%%%%%%%%%%%%%%%%%%%%%%%%%%%%%%%%%%
%%%%%%%%%%%%%%%%%%%%%%%%%%%%%%%%%%%%%%%%%%%%%%%%%%%%%%%%%%%%%%%%%%%%%%%%%
%                                                                       %
%   BEGIN OF FIGURES                                                    
%                                                                       %
%%%%%%%%%%%%%%%%%%%%%%%%%%%%%%%%%%%%%%%%%%%%%%%%%%%%%%%%%%%%%%%%%%%%%%%%%
\newpage
\begin{figure}
%\begin{center}
\leavevmode
\epsfxsize = 12cm
\epsffile[60 85 430 750]{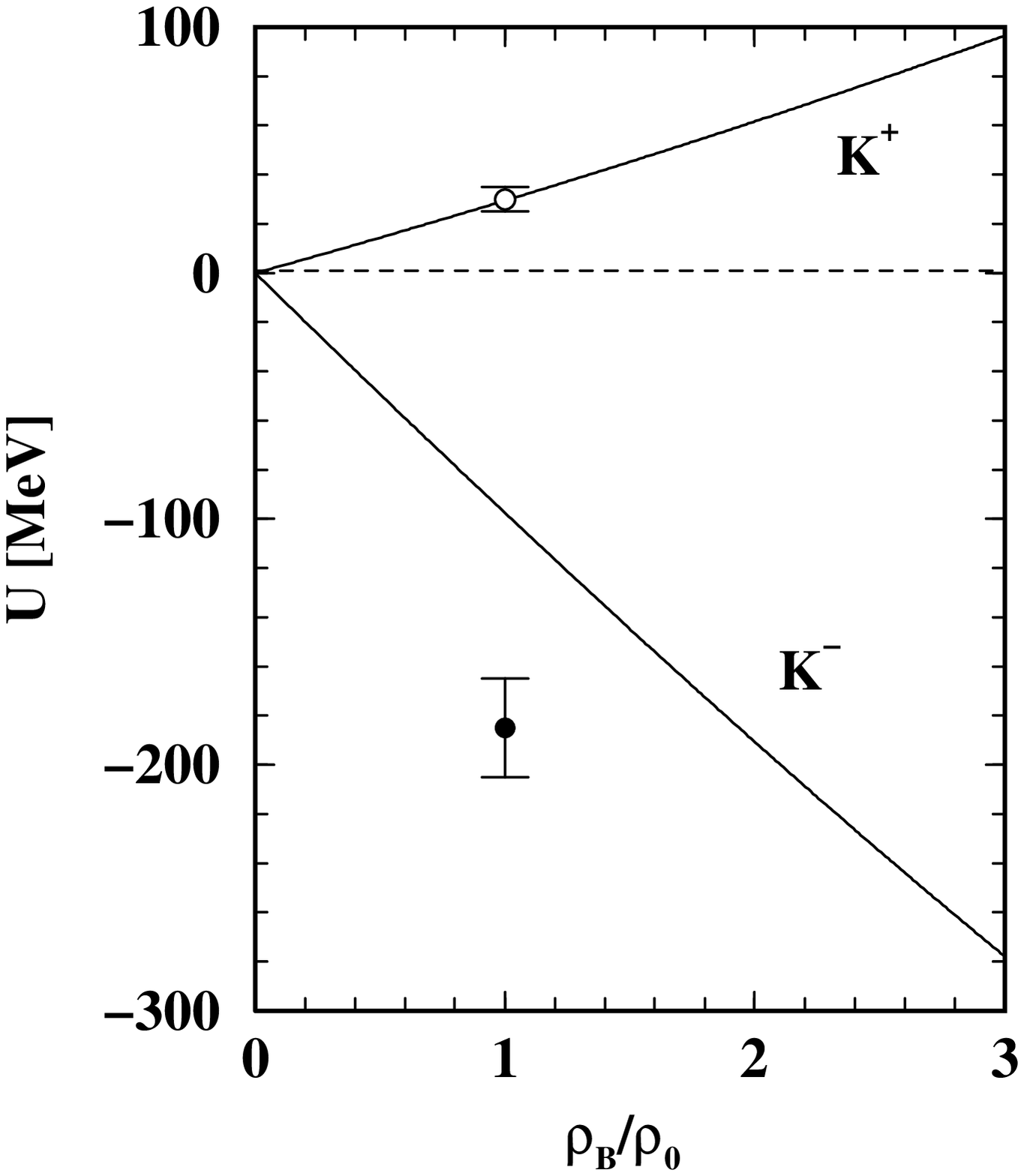}
%\end{center}
\caption{
}
\label{fig1}
\end{figure}
%%%%%%%%%%%%%%%%%%%%%%%%%%%%%%%%%%%%%%%%%%%%%%%%%%%%%%%%%%%%%%%%%%%%%%%%%
\newpage
\begin{figure}
%\begin{center}
\leavevmode
\epsfxsize = 12cm
\epsffile[20 90 460 660]{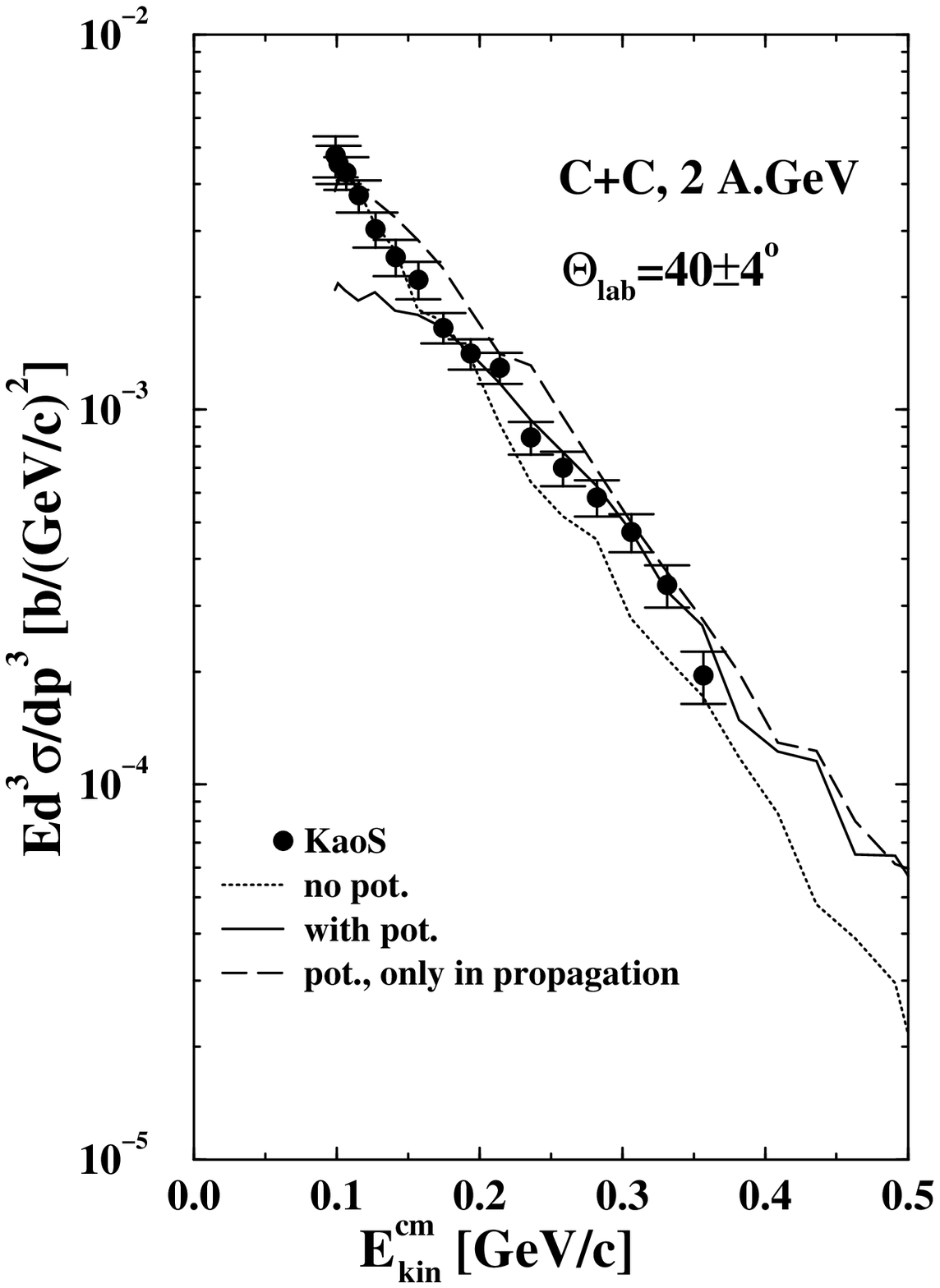}
%\end{center}
\caption{
}
\label{fig2}
\end{figure}
%%%%%%%%%%%%%%%%%%%%%%%%%%%%%%%%%%%%%%%%%%%%%%%%%%%%%%%%%%%%%%%%%%%%%%%%%
\newpage
\begin{figure}
%\begin{center}
\leavevmode
\epsfxsize = 12cm
\epsffile[60 75 430 750]{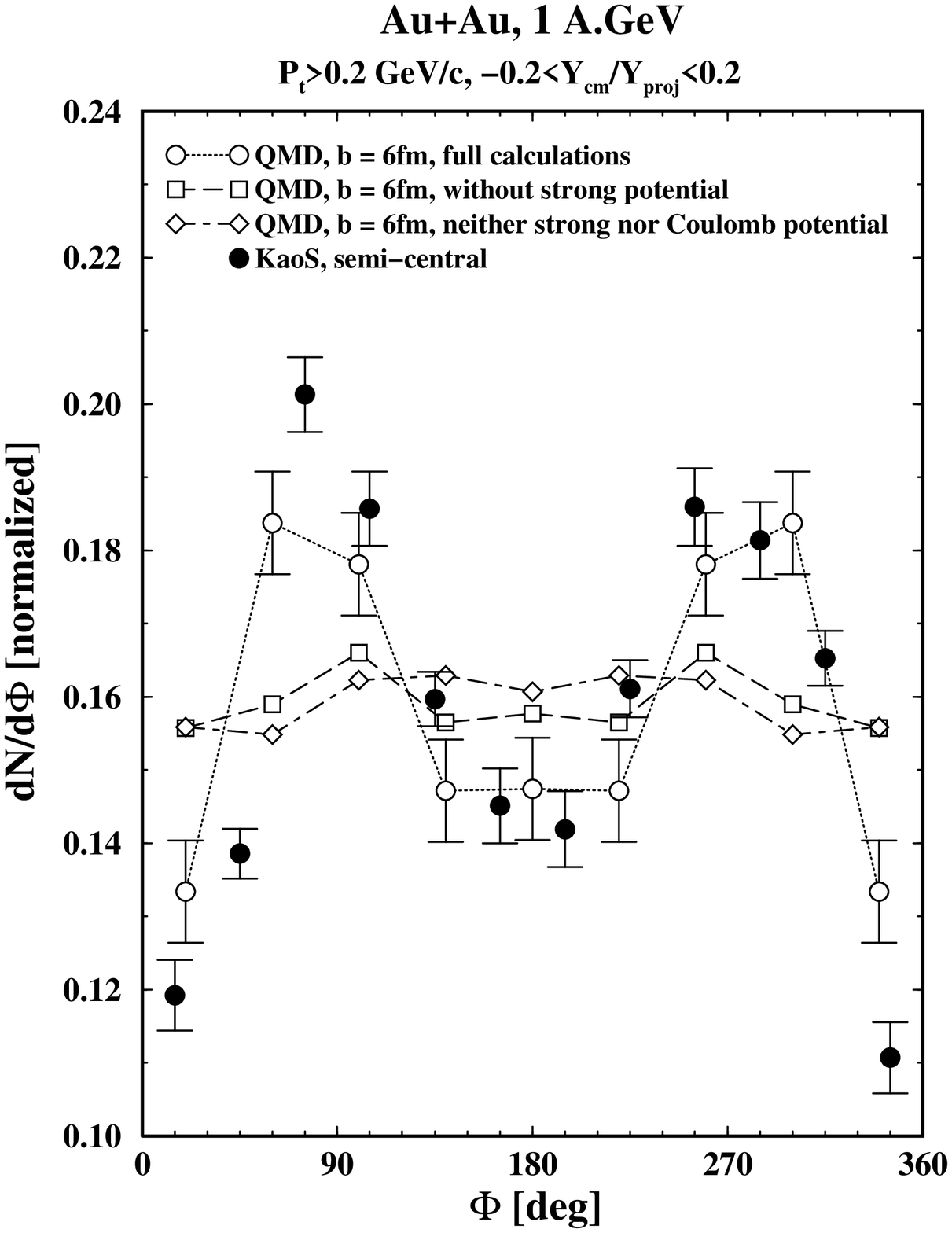}
%\end{center}
\caption{
}
\label{fig3}
\end{figure}
%%%%%%%%%%%%%%%%%%%%%%%%%%%%%%%%%%%%%%%%%%%%%%%%%%%%%%%%%%%%%%%%%%%%%%%%%
\newpage
\begin{figure}
%\begin{center}
\leavevmode
\epsfxsize = 12cm
\epsffile[60 70 430 750]{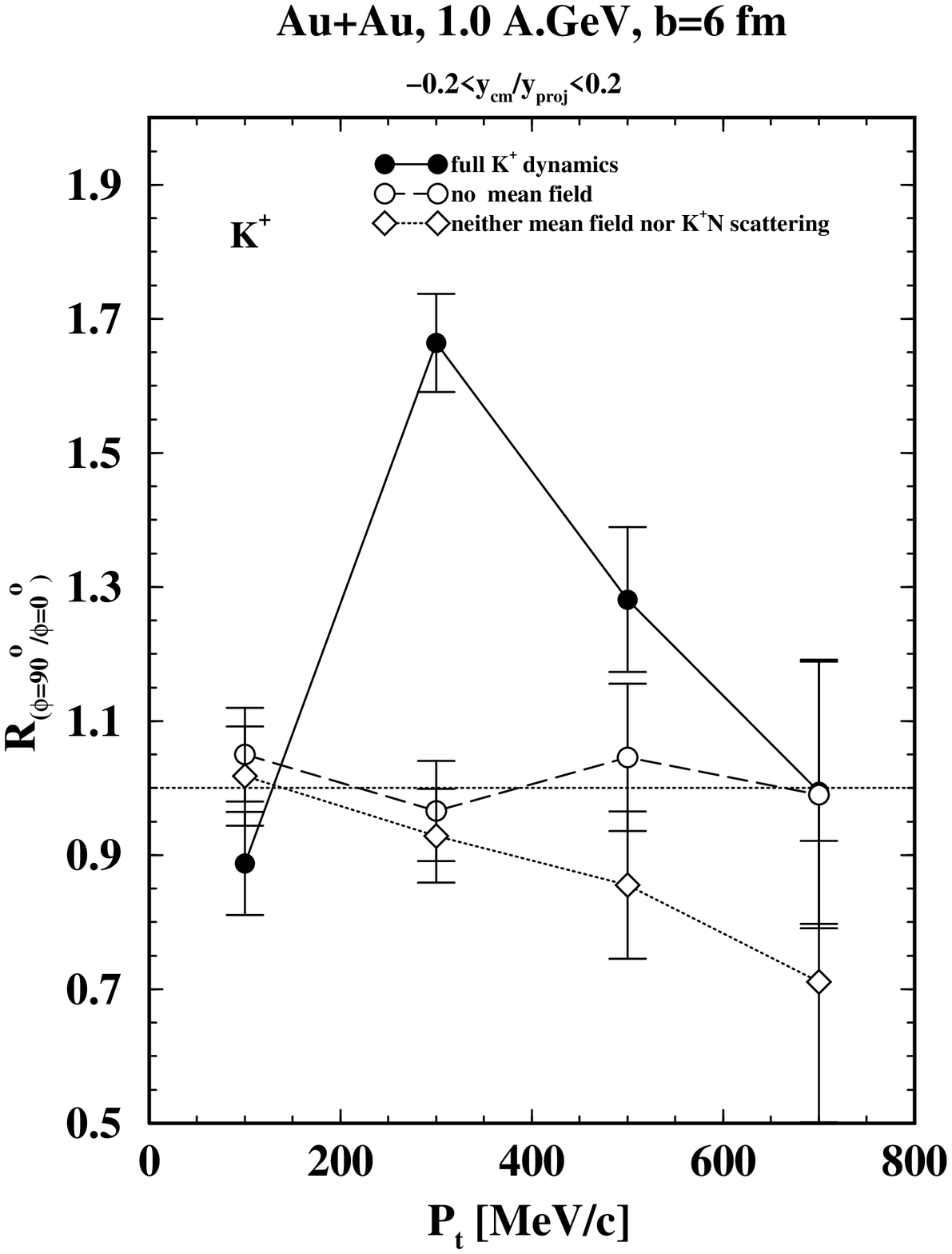}
%\end{center}
\caption{
}
\label{fig4}
\end{figure}
%%%%%%%%%%%%%%%%%%%%%%%%%%%%%%%%%%%%%%%%%%%%%%%%%%%%%%%%%%%%%%%%%%%%%%%%%
\newpage
\begin{figure}
%\begin{center}
\leavevmode
\epsfxsize = 12cm
\epsffile[60 75 430 750]{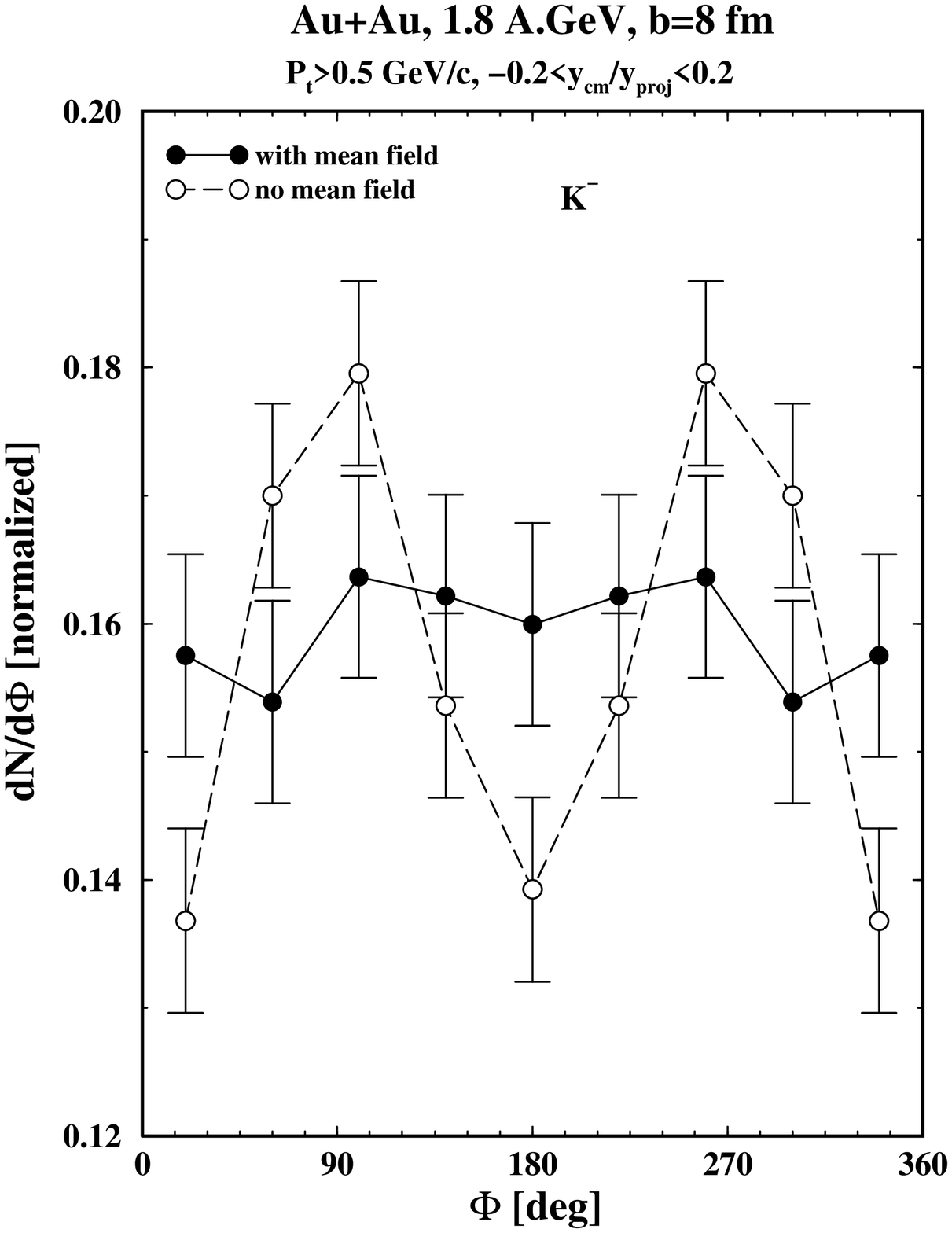}
%\end{center}
\caption{
}
\label{fig5}
\end{figure}
%%%%%%%%%%%%%%%%%%%%%%%%%%%%%%%%%%%%%%%%%%%%%%%%%%%%%%%%%%%%%%%%%%%%%%%%%
\newpage
\begin{figure}
%\begin{center}
\leavevmode
\epsfxsize = 12cm
\epsffile[60 70 430 750]{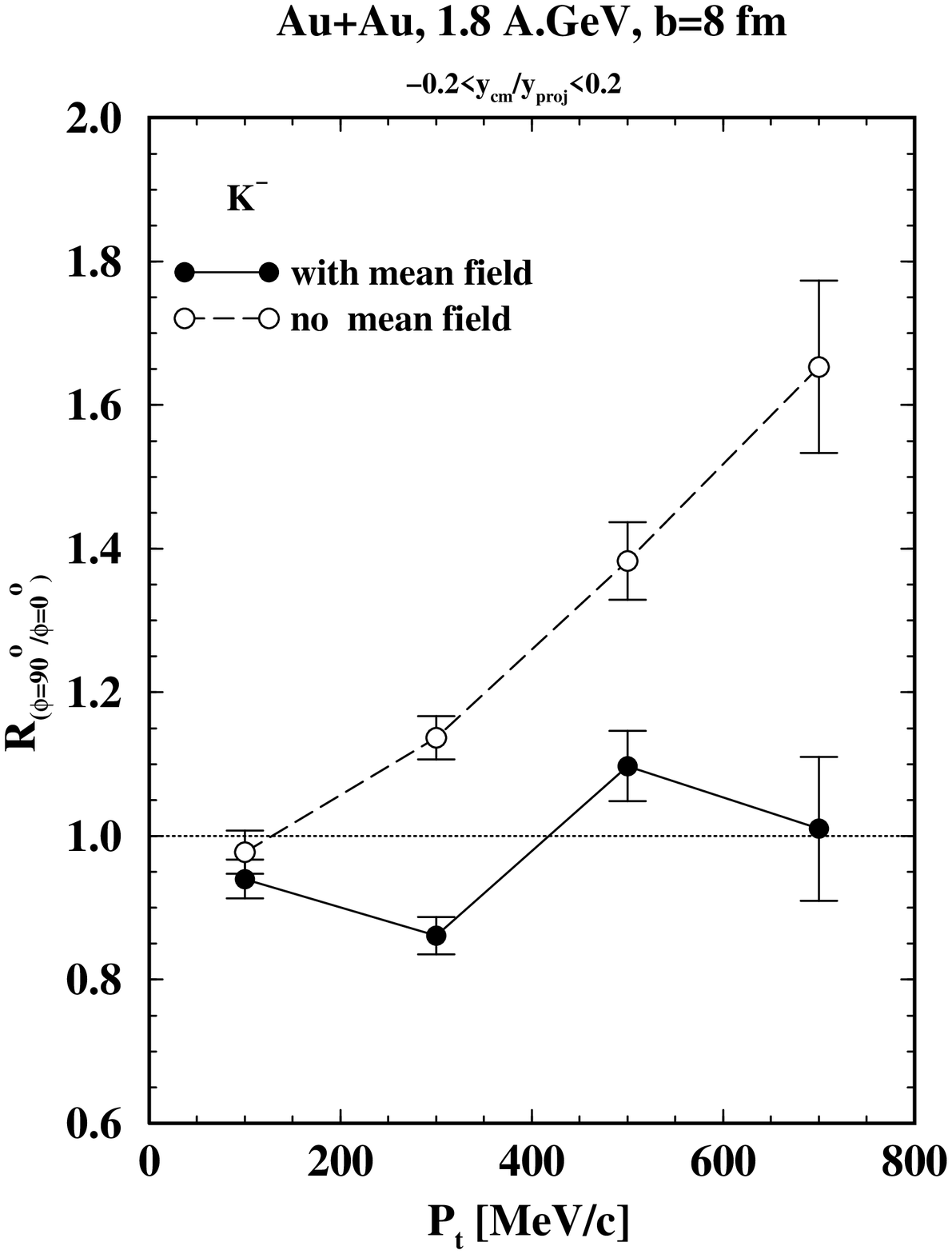}
%\end{center}
\caption{
}
\label{fig6}
\end{figure}
%%%%%%%%%%%%%%%%%%%%%%%%%%%%%%%%%%%%%%%%%%%%%%%%%%%%%%%%%%%%%%%%%%%%%%%%%
\newpage
\begin{figure}
%\begin{center}
\leavevmode
\epsfxsize = 12cm
\epsffile[60 75 430 750]{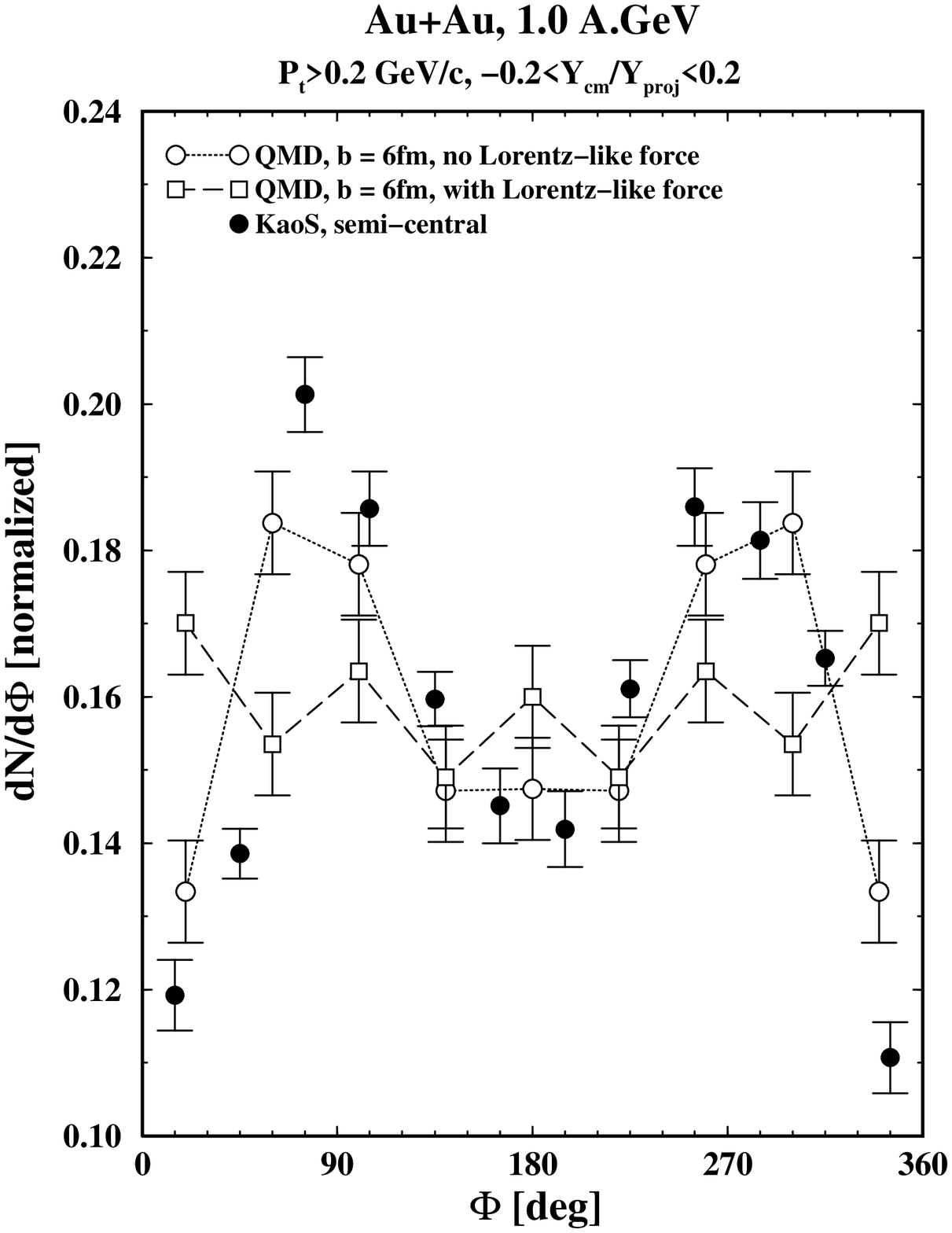}
%\end{center}
\caption{
}
\label{fig7}
\end{figure}
%%%%%%%%%%%%%%%%%%%%%%%%%%%%%%%%%%%%%%%%%%%%%%%%%%%%%%%%%%%%%%%%%%%%%%%%%
%                                                                       %
%%%%%%%%%%%%%%%%%%%%%%%%%%%%%%%%%%%%%%%%%%%%%%%%%%%%%%%%%%%%%%%%%%%%%%%%%
%                                                                       %
%%%%%%%%%%%%%%%%%%%%%%%%%%%%%%%%%%%%%%%%%%%%%%%%%%%%%%%%%%%%%%%%%%%%%%%%%
%                                                                       %
%   END OF FIGURES                                                      %
%                                                                       %
%%%%%%%%%%%%%%%%%%%%%%%%%%%%%%%%%%%%%%%%%%%%%%%%%%%%%%%%%%%%%%%%%%%%%%%%%
%%%%%%%%%%%%%%%%%%%%%%%%%%%%%%%%%%%%%%%%%%%%%%%%%%%%%%%%%%%%%%%%%%%%%%%
%%%%%%%%%%%%%%%%%%%%%%%%%%%%%%%%%%%%%%%%%%%%%%%%%%%%%%%%%%%%%%
\end{document}